\documentclass{article}
\usepackage[english]{babel}
\usepackage{graphicx}
\usepackage{amsmath}
\usepackage{natbib}
\usepackage{authblk}
\title{Competition is the underlying mechanism controlling viscous
  fingering and wormhole growth.}
\author[1,2,3]{Yoar Cabeza}
\author[2,3]{Juan J. Hidalgo}
\author[1,3]{Jesus Carrera}

\affil[1]{Institute of Environmental Assessment and Water
  (IDAEA), Spanish National Research Council (CSIC), Barcelona,
  Spain.}

\affil[2]{Universitat Polit{\`e}cnica de Catalunya (UPC),
Barcelona, Spain.}

\affil[3]{Associated Unit: Hydrogeology Group (UPC-CSIC)}
\date{}
\begin{document}
\maketitle
%
%
%
%
\begin{abstract}
  Viscous fingering and wormhole growth are complex nonlinear unstable
  phenomena. We view both as the result of competition for water in
  which the capacity of an instability to grow depends on its ability
  to carry water. We derive empirical solutions to quantify the
  finger/wormhole flow rate in single-, two-, and multiple-finger
  systems. We use these solutions to show that fingering and wormhole
  patterns are a deterministic result of competition. For wormhole
  growth, controlled by dissolution, we solve reactive transport
  analytically within each wormhole to compute dissolution at the
  wormhole walls and tip. The generated patterns (both for viscous
  fingering and wormhole growth under moderate Damk\"ohler values)
  follow a power law decay of the number of fingers/wormholes with
  depth with an exponent of -1 consistent with field observations.
\end{abstract}
%
%
\section{Introduction}
Wormhole growth and viscous fingering are unstable growth phenomena
controlled by interface dynamics~\citep{Szymczak2011}. Viscous
fingering occurs when a less viscous fluid displaces another one,
destabilizing the displacement front~\citep{Homsy1987}. Wormholes are
highly conductive flow channels generated by
dissolution~\citep{Fredd1998} that enhance the effective permeability
of the medium \citep{Cheung2002} and, under the appropriate boundary
conditions, the resulting flow rate \citep{Kaufmann2000,
  dreybrodt2005}. Understanding fingering patterns growth is of
interest for CO$_{2}$ storage where the caprock may be compromised due
to dissolution \citep{Luquot2009}, oil industry where it is a relevant
for well stimulation~\citep{Fredd1998, Golfier2002, Cohen2008},
enhanced oil recovery, and karst aquifers studies where fracture or
matrix dissolution can evolve from an initial wormhole
pattern~\citep{Hanna1998, Szymczak2011}.

In general, wormhole growth is controlled by flow, transport, and
chemistry~\citep{Singurindy2003, Edery2011, Szymczak2011,
  Hidalgo2015}. Dissolution causes permeability to increase locally,
which enhances the flow of aggressive water and promotes further
dissolution. The formation and shape of wormhole patterns is
controlled by the Damk\"{o}hler ($\text{Da}$) and P{\'e}clet
($\text{Pe}$) numbers~\citep{Fredd1998, Golfier2004,
  Golfier2006}. Dominant wormholes are characteristic of intermediate
water fluxes and $\text{Da}/\text{Pe}<1$ in which the aggressive water
flows into the biggest pores to form flow
channels~\citep{Golfier2004}. Conical wormholes generate when
$\text{Da}/\text{Pe} > 1$ and transport is dominated by diffusion but
instabilities can still grow. Under these conditions, the reactant
erodes the walls of the flow channels to form conical-shaped
wormholes. In contrast, viscous fingers can be regarded as
non-reactive wormholes ($\text{Da} = 0$) where growth occurs only at
the tip. Chemical reactions may also alter the rheology of the fluid
leading to viscous instabilities \citep{Nagatsu2007, Bunton2017}.

The growth pattern can be viewed as a flow problem. As one wormhole
grows, its flow rate increases and water flux reduces
elsewhere. Therefore, wormholes compete for fluid flow
\citep{Szymczak2006}. Competition and flow focusing are determinant
for the final pattern of non-reactive finger growth
\citep{Pecelerowicz2014} and the dissolution of fractures in the
genesis of karst \citep{Hanna1998, Szymczak2011}. The important role
of competition is also suggested by the observed decay of the number
of wormholes with depth, which follows a power law with exponent close
to -1~\citep{krug1993, budek2012, Upadhyay2015}.

The main conjecture of this work is that competition for flow is
responsible of the resulting viscous finger and wormhole patterns,
which means that the instability pattern is deterministic and
controlled by the flow redistribution. To test this conjecture, we
develop an empirical model that quantifies the competition. This model
is used to simulate viscous fingering and, when coupled to a reactive
transport model, wormhole growth. We compare the results for the final
finger-like pattern to the final wormhole pattern and evaluate how
competition controls both phenomena.
%
%
\section{Competition for flow model}
We develop a competition model based on the flow rate that a finger or
a wormhole (fingers for brevity) can carry in the presence of
others. The multiple-finger competition model is based on the single-
and two-finger interaction cases, which we examine first.

Fingers can be viewed as thin structures whose permeability is
infinite with respect to the displaced fluid (or matrix in the case of
wormholes), so that head is constant along them, thus leading to an
increasing flow towards its tip~\citep{Nilson1990, Daccord1993}
(Figure~\ref{fig:CaptureAreas}a). The finger's advance is proportional
to the flow rate it carries. This is approximately true as well for
wormhole growth because the dissolution capacity at the tip (volume of
rock that can be dissolved for the wormhole to advance) is also
proportional to the flow rate, at least for moderate dissolution
rates.
 
We solve the flow problem assuming that fingers grow as straight lines
in a two dimensional rectangular domain subject to a natural flux
(flow rate per unit width) $q_{N}$. Since permeability in the
dissolved area is much bigger than in the matrix, we can consider that
head is constant at the inflow boundary and along the fingers, so that
the natural flow rate is redistributed and concentrated along the path
of least resistance. To save numerical effort the top boundary of our
model is set at the finger's root and a Dirichlet boundary condition
is prescribed. An outgoing flux is prescribed at the bottom boundary
so that water enters the system through the top (upper boundary and
wormholes) and flows downwards. Under these boundary conditions and
considering that flow is governed by Darcy's law
$\mathbf{q} = - T \boldsymbol{\nabla} h$, where $T$ is transmissivity
and $h$ is head, the problem is simulated by
\begin{align}
  \label{eq:flow}
  &\nabla^{2} h = 0\\ 
  &h(0 \leq x \leq L_{i}, y_{i}) =0  \\ 
  &h(0, y) = 0 \\
  \label{eq:flow2}
  &T \left. \frac{\partial h}{\partial x}\right|_{x=L_x} = -q_{N}
\end{align}
where $y_{i}$ and $L_{i}$ are the location and length of the $i$-th
finger, for $i = 1, \ldots, n_{f}$, with $n_{f}$ the number of
fingers.  To avoid boundary effects $L_{x}, L_{y} \gg L_{i}$.

The flow rate along the $i$-th finger is given by
\begin{align}
  \label{eq:WH-FlowRate}
  Q_{i} = \int_{0}^{L_{i}} |q_{y}(x, y^{+}_{i})|\mathrm{d}x + \int_{0}^{L_{i}} |q_{y}(x, y^{-}_{i})|\mathrm{d}x.
\end{align}
where $y^{\pm}_{i}$ indicate that $q(x,y)$ is evaluated at the sides
of the finger.

Equations~(\ref{eq:flow}) -- (\ref{eq:flow2}) were solved using the
finite element code TRANSIN \citep{medina2003geostatistical}. The mesh
was highly refined around the fingers tips until convergence to
minimize numerical errors associated to the singularity at that area.
%
%
\subsection{The single- and two-finger cases}\label{sec:2whModel}
The flow rate through a single finger is proportional to its length
\citep{Cabeza2014}. Simulations showed that the flow rate captured by
a finger without neighbors equals
\begin{align}
  \label{eq:Qs}
  Q_{s} = 2 Lq_{N},
\end{align}
where $Q_{s}$ is the flow rate and the subscript $s$ refers to the
single-finger case.

\begin{figure}[htp]
  \centering
  \includegraphics[width=1\textwidth]{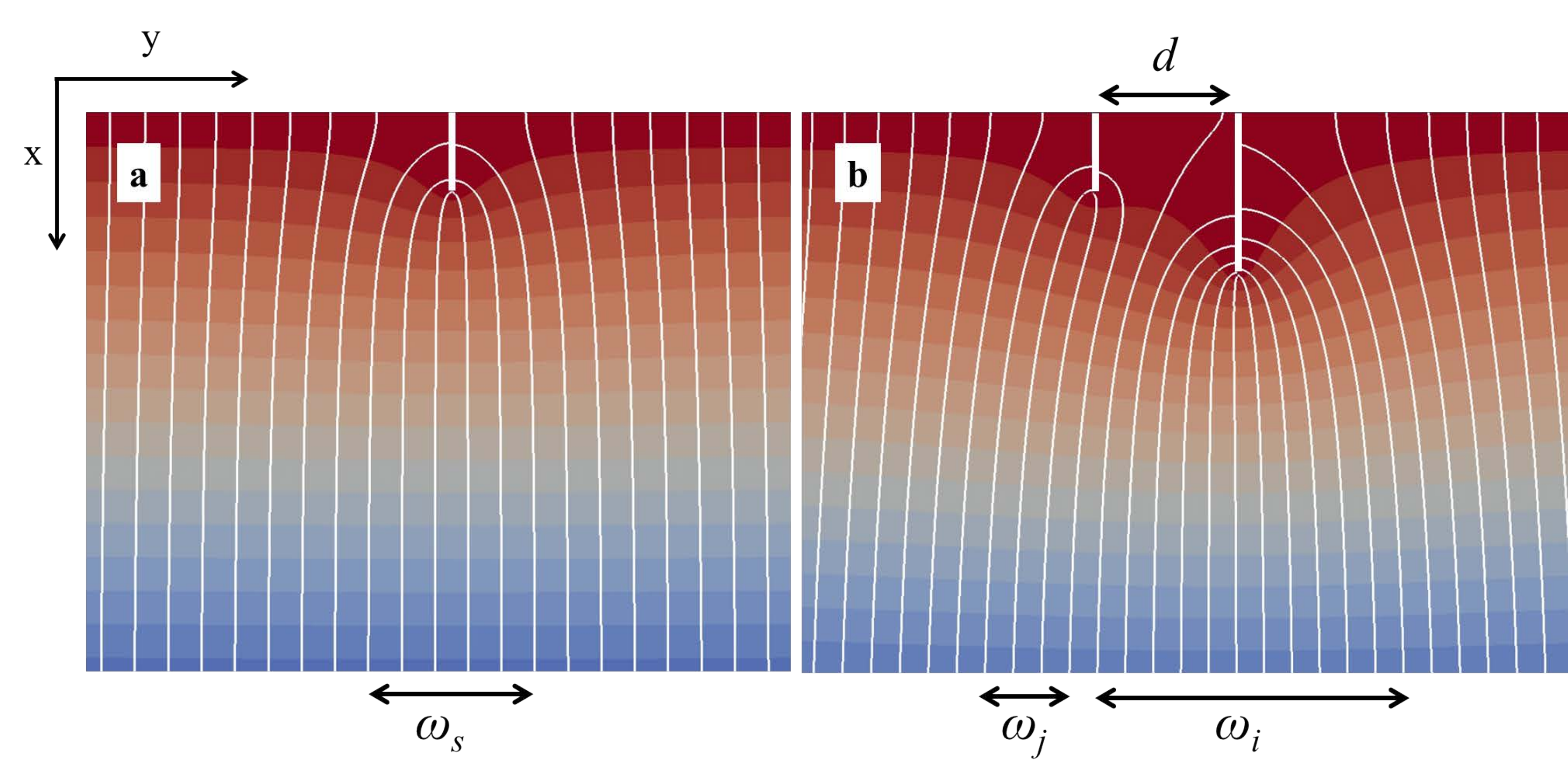}
  \caption{Flownet (colors for heads and white streamlines, water
    flows downwards) around (a) a single and (b) two competing
    fingers. The domain lateral boundaries (not shown) are far from
    the fingers to avoid boundary effects.  The streamlines that start
    at the root of each wormhole delimit the flow through the
    wormhole. The length between these streamlines at the bottom
    boundary far from the inflow is the capture area $\omega$ of each
    finger. Under competition the fingers screen each other and their
    capture areas are reduced with respect the single finger
    value. The reduction in the capture area is more notable for the
    short finger than for the long one.} \label{fig:CaptureAreas}
\end{figure}

By analogy to the concept of well catchment area used in hydrology, we
generalize this expression by defining the finger capture area
$\omega$ as the length that would carry the same flow rate as the
finger if traversed by the natural flux (i.e., flow rate of the
finger divided by $q_{N}$). Therefore, finger water flows across
area length $\omega$ at a long distance from the finger's
root~(Figure~\ref{fig:CaptureAreas}b). The capture area $\omega$ is a
natural length scale to characterize flow interference between
fingers. From \eqref{eq:Qs} the single finger capture area is
$\omega_{s} = 2L$. Note that $\omega$ has units of length in the 2D
competition model, because the thickness is implicit in the concept of
transmissivity.
 
When two fingers compete, they screen each other. Their capture areas
are smaller than those of single fingers and the flux across the
matrix in between becomes negligible
(Figure~\ref{fig:CaptureAreas}b). However the long finger captures
more water from the short one than vice versa. Screening increases
gradually during growth as the distance between fingers becomes small
compared to their length. Eventually, the long finger capture area
overlaps the short finger, which stops growing.  We define the
reduction of the $j$ finger capture area with respect to its single
finger capture area $\omega_{sj}$ caused by the presence of a neighbor
$i$ at a distance $d_{ij}$ as
\begin{align}
  \label{eq:CaptureAreaRed}
  \omega_{rji} = \omega_{sj} - \omega_{j}
\end{align}
where $\omega_{j} = \omega_{j}\left(L_i,L_j,d_{ij}\right)$ is the
actual capture area. The concept is illustrated in
Figure~\ref{fig:CaptureAreas}, which displays the single finger
capture area $\omega_{s}$ and the actual capture areas $\omega_{i}$
and $\omega_{j}$ ($L_{i} > L_{j}$) in a competition system as defined
by the water flowing out of the wormholes (delimited by the
streamlines exiting the finger). This flow rate is identical to the
one captured at the inlet, which we assume to be well connected.

To determine the reduction of the fingers capture areas on fingers
screened by a neighbor, we simulated the two-finger competition in a
dimensionless system with $T=1$, $q_{N}=1$, and wormholes lengths
$L_{i} \in [0.04, 1]$, $L_{j} \in [0.02, 0.2]$, separated a distance
$d \in [0.02, 0.2]$. Assuming $L_{i} \geq L_{j}$, the reduction of
both fingers capture areas can be reproduced by the following
empirical model (see supplementary material)

\begin{align}
  \label{eq:Reduction2Wh1}
  \omega_{rij} &= \beta \alpha L_{j} \\
  \label{eq:Reduction2Wh2}
  \omega_{rji} &= \omega_{sj} - \alpha L_{j},
\end{align}
where
\begin{align}
  \label{eq:alfa}
  \alpha = \min{\left\{2, \frac{\arctan(U)}{1-0.2 \arctan(U)}\right\}},
\end{align}
\begin{align}
  \label{eq:U}
  U = 0.0214 \frac{L_{j}}{d} + \frac{2d}{L_{i}-\beta L_{j}},
\end{align}
and
\begin{align}
  \label{eq:beta}
  \beta = \min{\left\{0.42\ln\left(1+\frac{L_{j}}{d}\right),
\frac{L_{j}}{L_{j}+2d}\right\}}
\end{align}

Comparing~\eqref{eq:CaptureAreaRed} and \eqref{eq:Reduction2Wh2}, we
see that $\omega_{j} = \alpha L_{j}$. That is, $\alpha$ can be viewed
as the dimensionless capture area of the short finger. The area from
which the long finger captures water \eqref{eq:Reduction2Wh1}
decreases proportionally to the shorter finger capture area while the
reduction for the short finger is stronger and depends fundamentally
on the difference of lengths and the distance between fingers. Note
that the reduction is zero when either $d \to \infty$ or $L_{j} = 0$.
%
%
\subsection{Multiple-finger competition}
We extend now the competition model to a multiple-finger system. Since
fingers are modeled as Dirichlet boundaries, the two-finger model
cannot be extended by superposition to determine
$\omega_{r}$. However, we can apply the superposition principle if the
impact of neighboring fingers is expressed in terms of their flow
rates using an approach similar to that of~\citet{Murdoch1994}. That
is, the reduction in the flow rate of the $i$-th finger in a system
with $n_{f}$ fingers is the result of the reductions caused by the
flow rates of each neighboring finger. We now recall that
$\omega_{rij}$ is the reduction in flow rate of the $i$-th finger
caused by a flow rate ${\omega_{sj}- \omega_{rji}}$ at the $j$-th
finger. Therefore, the total reduction in flow rate of the $i$-th
finger, caused by the actual flow rates, ${\omega_{j}}$, in the
neighboring fingers is be equal to
\begin{align}
  \label{eq:TotalReduction}
  \omega_{ri} =  \sum^{n_{f}}_{\substack{j=1 \\ j \neq i}} \frac{\omega_{rij}}{\omega_{sj}-\omega_{rji}} \omega_{j},
\end{align}
where $\omega_{j}$ is the capture area of the $j$-th finger (unknown
at this stage).  Since the total reduction in the capture area in the
$i$-th finger is $\omega_{ri} = \omega_{si} - \omega_{i}$, one can
finally write
\begin{align}
  \label{eq:TotalReduction2}
  \sum^{n_{f}}_{j=1} \left[\delta_{ij} + (1- \delta_{ij})\frac{\omega_{rij}}{\omega_{sj}-\omega_{rji}} \right] \omega_{j} = \omega_{si};  \,\, i=1,\ldots, n_{f},
\end{align}
where $\delta_{ij}$ is the Kronecker delta, and $\omega_{rij}$ and
$\omega_{rji}$ are computed using \eqref{eq:Reduction2Wh1} and
\eqref{eq:Reduction2Wh2} respectively. Therefore, the capture areas in
the multiple-finger system can be obtained solving the linear system
of equations~\eqref{eq:TotalReduction2} in which the coefficients are
given by the geometry of the fingers.
%
%
\section{Dissolution capacity in a wormhole}
The advance of a wormhole depends on the dissolution capacity at its
tip \citep{Szymczak2006}. Dissolution processes at the wormhole tip
are complex because the water flux field is singular at the tip;
dissolution does not occur at the wormhole wall, but in the matrix
around; and advection and diffusion both in the matrix and the
wormhole compete with dissolution, leading to a range of wormhole
morphologies. We discuss these processes in the supplementary
material. The dissolution capacity at the tip is given by the
difference between the dissolution capacity of the incoming water and
the dissolution at the wormhole walls (actually, not at the walls but
within the matrix around, yielding a high transmissivity dissolution
halo around the walls and tip
\citep{Wei1990,Szymczak2009}). Dissolution within the halo around the
walls can be approximated as if it occurred at the wormhole wall for
the conditions favoring wormhole growth (see supplementary material
). While the transmissivity within this halo can be high, we assume
that most of the flow concentrates within the wormhole. Wormhole
advance is slow compared to the flow and transport time scales and we
model it as a quasi-steady state process in which the dissolution
concentrates at the walls and the tip, as controlled by the transport
equation
\begin{align}
  \label{eq:transport}
  v(x)\frac{\partial c}{\partial x} = D \frac{\partial^2 c}{\partial y^2}.
\end{align}
In \eqref{eq:transport} $c = (C - C_{eq})/(C_{0} - C_{eq})$ with $C$
the concentration, $C_{eq}$ the equilibrium concentration, and $C_{0}$
the concentration at the inlet, $0 \leq x \leq L$ is the direction
along the wormhole growth, $-b(x) \leq y \leq b(x)$ is the direction
along the wormhole half-width $b(x)$, the flow velocity
$v(x) = Q/2b(x)$ or $\omega q_{N}/2b(x)$ in terms of the wormhole's
capture area, $D$ is the diffusion coefficient, and longitudinal
diffusion is considered negligible. Boundary conditions are
\begin{align}
  \label{eq:c-inlet}
   c(0,y) &= 1\\
  \label{eq:adv-disp_bound}
  \left. D\frac{\partial c(x,y)}{\partial y}\right|_{y=b(x)}&=-k_{f}c(x,b(x)),
\end{align}
where $k_{f}$ is an effective kinetic constant that depends on the
volumetric kinetic rate, pore diffusion, and specific surface in the
halo surrounding the wormhole (see supplementary material).

Note that this model assumes that all the captured water flows through
the entire length of the wormhole. In reality
(Figure~\ref{fig:CaptureAreas}), a portion flows laterally through the
lower part of the wormhole contributing to the dissolution halo and,
possibly, to the development of lateral wormholes or even wormhole
bifurcation.

Assuming that the variation of the wormhole half-width $b(x)$ with
depth is smooth, we have solved
\eqref{eq:transport}--\eqref{eq:adv-disp_bound} in a warped coordinate
system \citep{Ranz1979} (see supplementary material)
\begin{align}
  \label{eq:y2z}
  z=\frac{y}{b(x)}; \,\,  \tau =\frac{D}{\omega q_{N}} \int_{0}^{x} \frac{dx}{b(x)}.
\end{align}

In each growth time step $\Delta t$ the wormhole half-width increases
as
\begin{align}
  \label{eq:int:delta_width}
  \Delta b(\tau) = v_{d} k_{f} c(\tau, 1),
  \end{align}
where $v_{d} = V_{M}(C_{0}-C_{eq})\Delta t /(1-\phi)$ is the volume of
porous matrix dissolved per unit flow rate of water and unit
dimensionless concentration, with $V_{M}$ the molar volume and $\phi$
the matrix porosity. The advance of the wormhole is proportional
to the dissolution capacity at its tip $\overline{c}(\tau(L))$
\begin{align}
  \label{eq:int:delta_length}
\Delta L = \lambda \frac{q_{N} \omega v_{d}}{b}\left( \frac{1}{2} - \int_{0}^{\tau(L)} \text{Da}^{\ast}(\tau) \hat{c}(\tau,1) \, d\tau\right)  = \frac{q_{N} \omega v_{d}}{b} \overline{c}(\tau(L)),
\end{align}
where $\lambda$ is the fraction of the wormhole flow rate effectively
contributing to the wormhole advance, $b$ is the mean half width of
the wormhole (assumed constant in our analytical solution). This
solution depends only on the Damk{\"o}hler number
$\text{Da}^{\ast} = k_{f}b/D$ and
$\tau(L) = DL/\omega q_{N}b = 1/\text{Pe}$. In the case of a single
wormhole $\text{Pe} = 2q_{N}b/D$, which does not depend on the
wormhole length because $\omega = 2L$ from~(\ref{eq:Qs}). Note also
that the factor $\lambda$ affects the time scale for significant
wormhole advance, but it is of the order of magnitude of unity (around
0.3) as we argue in the supplementary material for tubular wormholes.

\begin{figure}[htp]
  \centering
 \includegraphics[width=0.9\textwidth]{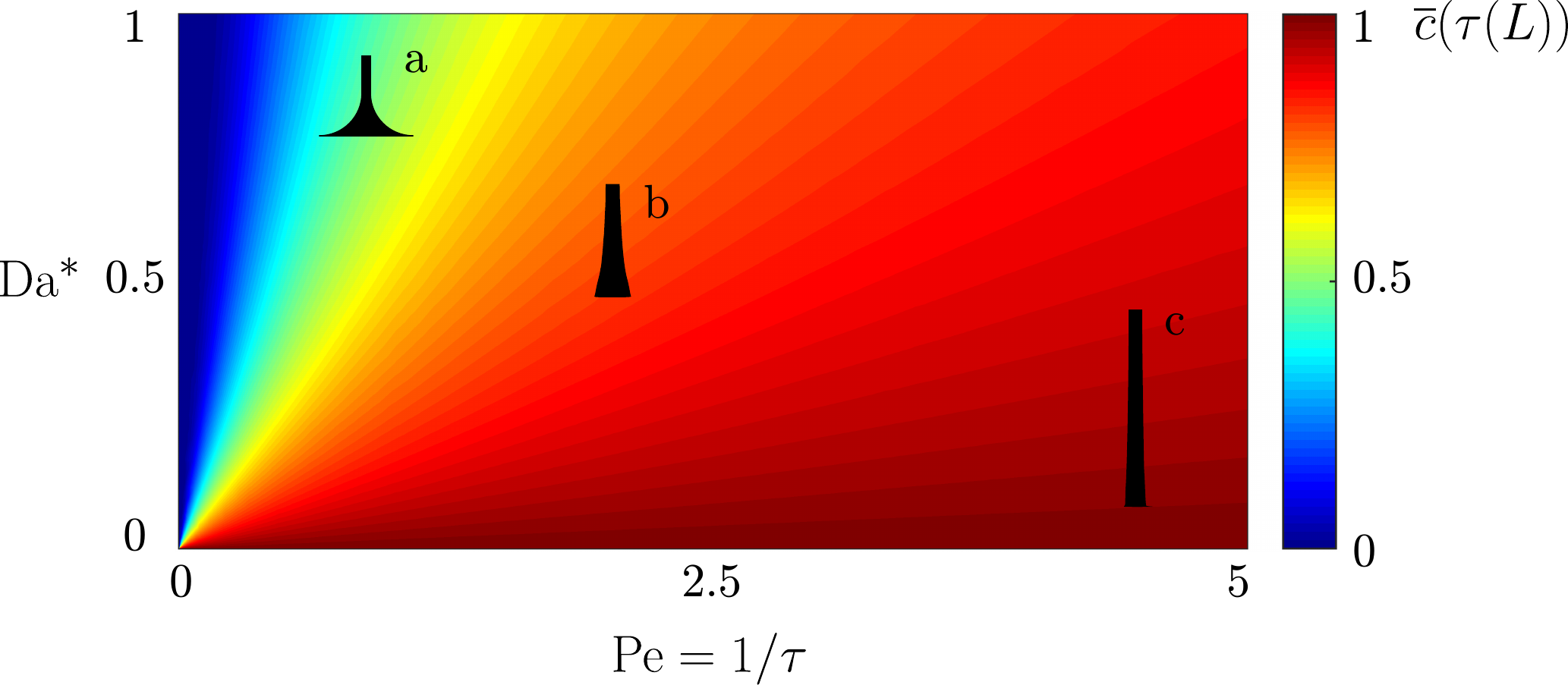}
 \caption{Average concentration at the tip of an initially straight
   wormhole for different values of $\text{Da}^{\ast}$ and
   $\text{Pe}$. The insets represent the expected wormhole shape under
   the obtained dissolution regime (the transverse scale is
   exaggerated). Under large $\text{Da}^{\ast}$ (a, b) the dissolution
   capacity is spent in widening the wormhole, which develops a
   conical shape. For comparable $\text{Pe}$ and $\text{Da}^{\ast}$
   (c) the dissolution capacity is almost intact at the tip, so the
   wormhole grows straight without widening at the walls. For very
   large $\text{Da}^{\ast}$ and/or very small $\text{Pe}$, dissolution
   occurs along a flat surfaced and wormholes do not
   develop. Reversely for very small $\text{Da}^{\ast}$ and/or very
   large $\text{Pe}$, dissolution occurs within the matrix and
   wormholes do not develop either }\label{fig:Collapse}
\end{figure}

Figure~\ref{fig:Collapse} displays the dissolution capacity at the tip
of an initially straight wormhole of half-width $b_{0}$. The
Damk\"{o}hler number $\text{Da}^{\ast}=k_{f}b_{0}/D$ is the only
controlling parameter because $\text{Pe} = 1/\tau(L)$ is actually the
dimensionless wormhole length. For low $\text{Da}^{\ast}$ the
dissolution capacity concentrates at the wormhole tip so that it grows
straight. For large $\text{Da}^{\ast}$ dissolution at the wall becomes
important and the wormholes develop a conical shape in agreement with
the results of \cite{Golfier2004}, where $\text{Pe} \times \text{Da}$
is equivalent to $\text{Da}^{\ast}$ here.
%
%
\section{Results}
We validate the competition model against the viscous fingers
experimental data of \cite{Pecelerowicz2014} and the wormhole
numerical data from P. Szymczak (personal communication). We start the
simulation using the measured initial lengths and separations. The
impervious lateral boundaries of the viscous fingers experimental
setup and the periodic boundary conditions of the wormholes numerical
simulations are simulated using the method of images since the
competition model was derived for an infinite medium. The capture
areas are computed every time step using \eqref{eq:TotalReduction2}
and the lengths updated according to \eqref{eq:int:delta_length}. When
the capture area of a finger or wormhole becomes zero, it is
considered inactive and removed from the system. The simulation stops
when only one finger or wormhole remains active. Note that the viscous
finger case is equivalent to a particular wormhole growth case with
$\text{Da}^{\ast} = 0$ so that the fingers grow proportionally to
their flow rate.

\begin{figure}[htp]
  \centering
  \includegraphics[width=0.9\textwidth]{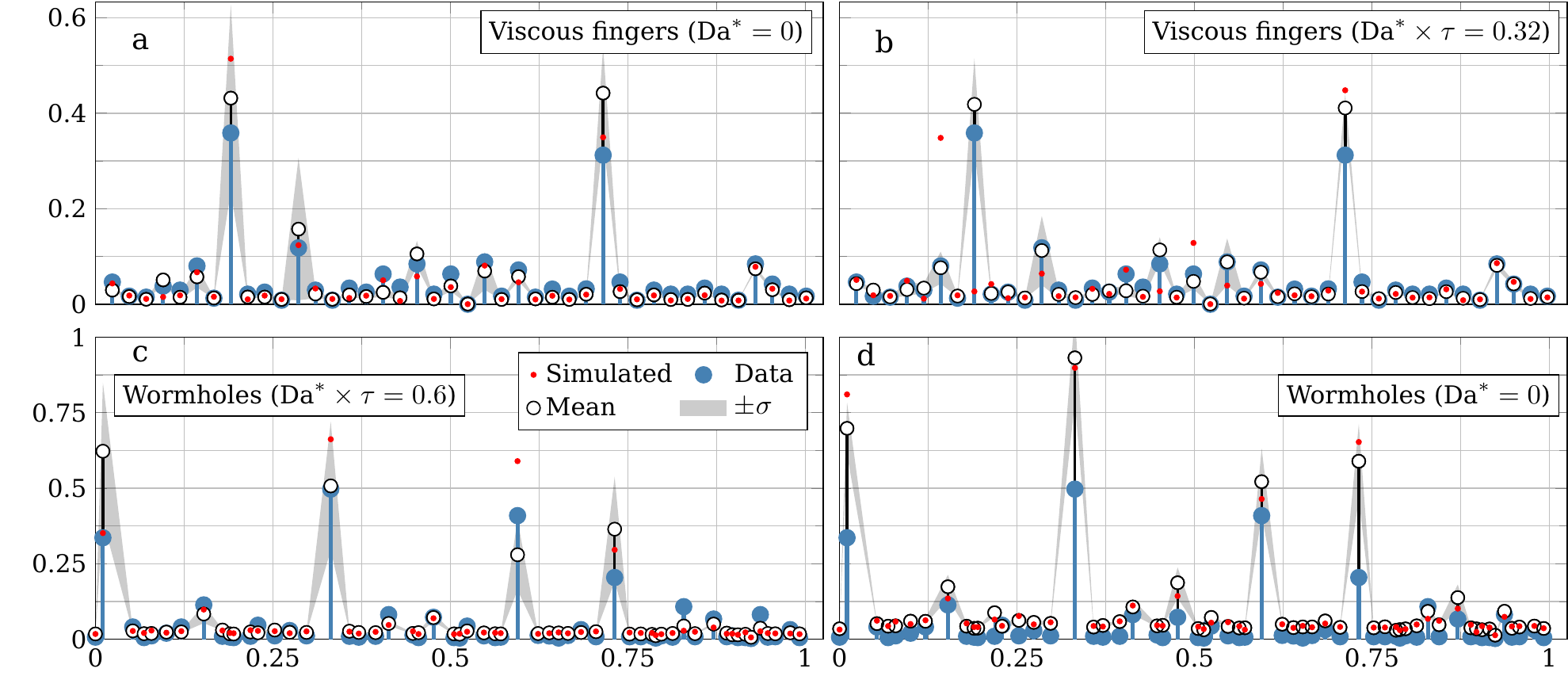}
  \caption{Distribution of viscous fingers (a, $\text{Da}^{\ast} =0$)
    and wormholes (c, $\text{Da}^{\ast} \times \tau = 0.6$) growing
    patterns with initial perturbations extracted from
    \cite{Pecelerowicz2014} and P. Szymczak (personal communication)
    respectively.  The blue dots are the measured lengths, the white
    dots are the lengths computed with the measured initial lengths,
    the red dots represent the mean value of lengths obtained for 2000
    realizations by perturbing randomly the measured initial lengths,
    and the gray area delimits two standard deviations around the
    mean. The number of surviving fingers fingers capable to take
    water from their surroundings and grow at the expense of the
    others is reduced as competition develops. To test the importance
    of competition the same simulations were repeated adding chemical
    reactions to the viscous finger system (b,
    $\text{Da}^{\ast} \times \tau = 0.32$) and removing them from the
    wormhole system (d, $\text{Da}^{\ast} =0$). Chemical reactions
    decrease the screening capacity of the fingers, which results in
    shorter lengths. The opposite effect is observed for
    wormholes. However, the mean final pattern is very similar in both
    cases which suggest that competition is the main controlling
    mechanism. }\label{fig:Pattern1}
\end{figure}

Figures~\ref{fig:Pattern1}a and \ref{fig:Pattern1}c show that the
competition model reproduces correctly the experimental and numerical
patterns (the correlation coefficient varies between 0.96 and
0.965). Differences can be attributed to uncertainty in the initial
conditions and to the approximations in the empirical model. We
performed a sensitivity analysis to the initial conditions by adding a
random perturbation to the initial lengths of the order of the
measurement error. The results of the 2000 realizations show that the
winning fingers are not affected (Figure~\ref{fig:Pattern1}a), which
confirms the deterministic behavior of the competition model. Only
small differences are observed in the moment when the shortest fingers
are completely screened.

The competition model tends to overestimate the length of the
wormholes (Figure~\ref{fig:Pattern1}c). Numerical data at the last
available time shows that there still are two active wormholes,
although there is a clear winner. The competition model does not stop
until there is only one active wormhole, therefore, prolonging the
growth of the longest wormholes. Growth is also faster in the
competition model because under the assumption of wormholes being
smaller than the domain size, there is no water limitation and long
wormholes speed up as the capture water from screened ones. This
differs with the numerical observations in which wormholes grow at a
constant speed during a time interval. Differences in growth speed can
contribute to the overall overestimation of the final lengths of the
pattern. However, relative wormhole's lengths and pattern geometry
remain unaffected.

To assess the effect of chemistry we repeated the simulations with the
same initial conditions but adding dissolution to the viscous finger
system (Figure~\ref{fig:Pattern1}b,
$\text{Da}^{\ast} \times \tau = 0.32$) and removing it from the
wormhole system (Figure~\ref{fig:Pattern1}d,
$\text{Da}^{\ast} \times \tau = 0$). We can see how the chemistry
alters slightly the final pattern. Dissolution affects mainly the
geometry of the secondary fingers (Figure~\ref{fig:Pattern1}b), which
advance relatively less and at a lower velocity than in the
non-reactive case because a reduced flow rate implies that an
increasing fraction of the dissolution capacity is used at the
walls. As a results the screening capacity of long wormholes is
reduced, which temporarily increases the flow through the secondary
wormholes. This reflected in an increasing range of finger length. The
opposite happens when chemistry is removed from the wormhole
system. The lengths increase and total screening occurs
earlier. Although the systems are intrinsically unstable and small
changes in the initial perturbation or in the growth rate affect the
final pattern and may lead to different winning fingers, the mean
length values remain stable. This supports the conjecture that
competition controls the growth. 

\begin{figure}[htp]
  \centering
  \includegraphics[width=1\textwidth]{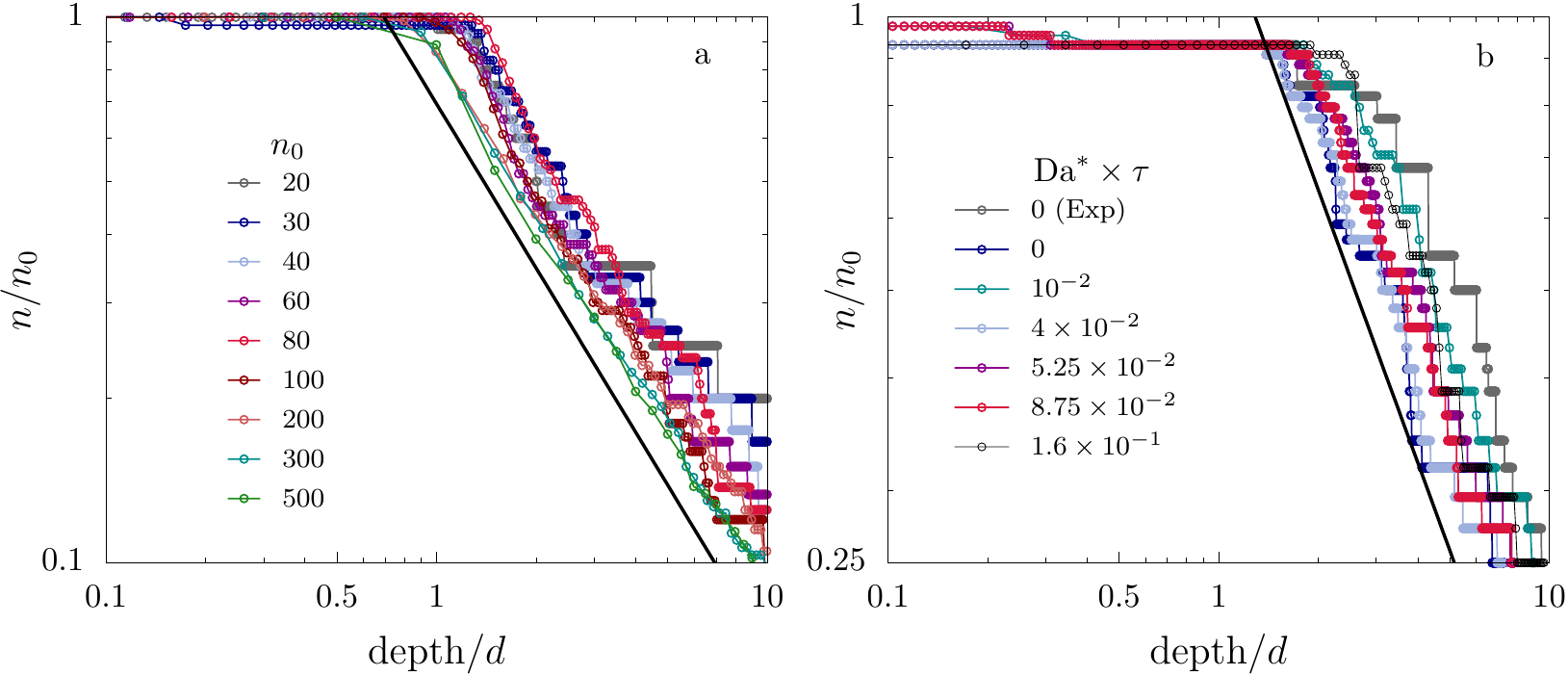}
  \caption{Evolution of the number of viscous fingers $n$ normalized
    by the initial number of fingers $n_{0}$ versus depth for
    simulations with $20\leq n_{0}\leq 500$  and initial lengths between
    $10^{-4}$ and $10^{-1}$ (a). The decay of $n/n_{0}$ follows a
    power law with an exponent close to -1 once the fingers enter the
    competition regime. This means that fingering patterns grow in a
    self similar way independently of the number of fingers, the
    distances between them, and the initial perturbation. The
    evolution of wormhole patterns under the effect of chemical
    reactions (b) displays a similar scaling.  The same initial
    lengths and $\text{Da}^{\ast} \times \tau$ values as in
    figure~\ref{fig:Pattern1} were used.}\label{fig:PowerLaw}
\end{figure}

Finally we simulate finger and wormhole growth in systems with
$20 \leq n_{0} \leq 500$ equidistant perturbations with random initial
lengths uniformly distributed in $(0, 0.1d]$ with $d = 1/(n_{0} -1)$
in order to keep every finger far from the competition regime at early
times. Results (Figure~\ref{fig:PowerLaw}a) shows that, independently
of the initial perturbation, the density of fingers $n/n_{0}$, where
$n$ is the number of active fingers, is constant at shallow depths,
which corresponds to the growth at full capacity of the
fingers. Competition starts when the length of the fingers is of the
order of the initial separation $d$ and the fingers capture areas
overlap. Then the number of surviving fingers decreases and
$n/n_{0} \sim 1/\text{depth}$ independently of the initial
configuration. This means that, if the ratio $n_{0}$/$L_{y}$ is kept
constant, the same self-similar pattern will emerge regardless the
size of the system. These results imply that competition promotes a
universal deterministic pattern of viscous fingers growth at any
scale.

Chemistry (Figure~\ref{fig:PowerLaw}b) does not alter the overall
scaling of the wormhole competition and the same dependence of
$n/n_{0}$ with depth is observed for moderate
$\text{Da}^{\ast} \times \tau$ values. The product
$\text{Da}^{\ast} \times \tau$ gives us the mean ratio of dissolution
power versus flow rate in the system.

%
%
\section{Conclusions}
The main conjecture of this work was that viscous fingering and
wormhole patterns are deterministic and defined by the redistribution
of flow driven by the fingers or wormholes geometry as they grow. We
developed a competition model based exclusively on the calculation of
the fingers flow-carrying capacities, which depends on distance to
their neighbors. When coupled to the advection-diffusion-reaction
equation, this model also reproduces wormhole growth. The results for
multiple-finger/wormhole simulations show that a deterministic
self-organized pattern, the number of fingers decreases linearly with
depth, emerges once the instabilities start competing for water. This
behavior is similar to the one observed in numerical simulations,
experiments, and in the field. We conclude that the flow-capturing
effect explains the final pattern and controls in a deterministic way
the distribution and density of fingers or wormholes in the system.

\paragraph{Acknowledgments}
Data used for producing the figures can be obtained by solving the
respective equations given in the supplementary material. We thank
Prof. Szymczak for the helpful discussion and providing the wormhole
data. JJH acknowledges the support of the European Research Council
through the project MHetScale (FP7-IDEAS-ERC-617511) and the Spanish
Ministry of Ministry of Science, Innovation and Universities ``Ram\'on
y Cajal'' fellowship (RYC-2017-22300). YC and JC acknowledge funding
Horizon 2020 EU project ACWAPUR (645782). JC acknowledges funding by
the Spanish Ministry of Economy and Competitiveness project
MEDISTRAES-II (CGL2013-48869-C2-2-R).
%
%
%
%

%
%
\end{document}


\maketitle
\noindent\textbf{Contents of this file}
\begin{enumerate}
\item Text S1
\end{enumerate}
\noindent\textbf{Introduction}
In this supplementary material we discuss (1) how the empirical model
of wormhole and viscous fingers competition fits the numerical
simulation data; (2) the analytical solution for the steady state
concentration distribution along a wormhole with dissolution around
the walls and tip, and; (3) the impact of a dissolution halo around
the wormhole.

\noindent\textbf{Text S1.}

\section{The single- and two-fingers cases}
We performed flow simulations with one and two fingers in a
dimensionless $2\times 2$ domain with unit transmissivity and unit
head gradient between the top and bottom boundaries (equations (1) -- (4)
in the manuscript). All wormholes longitudes were scaled with the
characteristic domain height.

\begin{figure}
\centering
\includegraphics[width=0.65\textwidth]{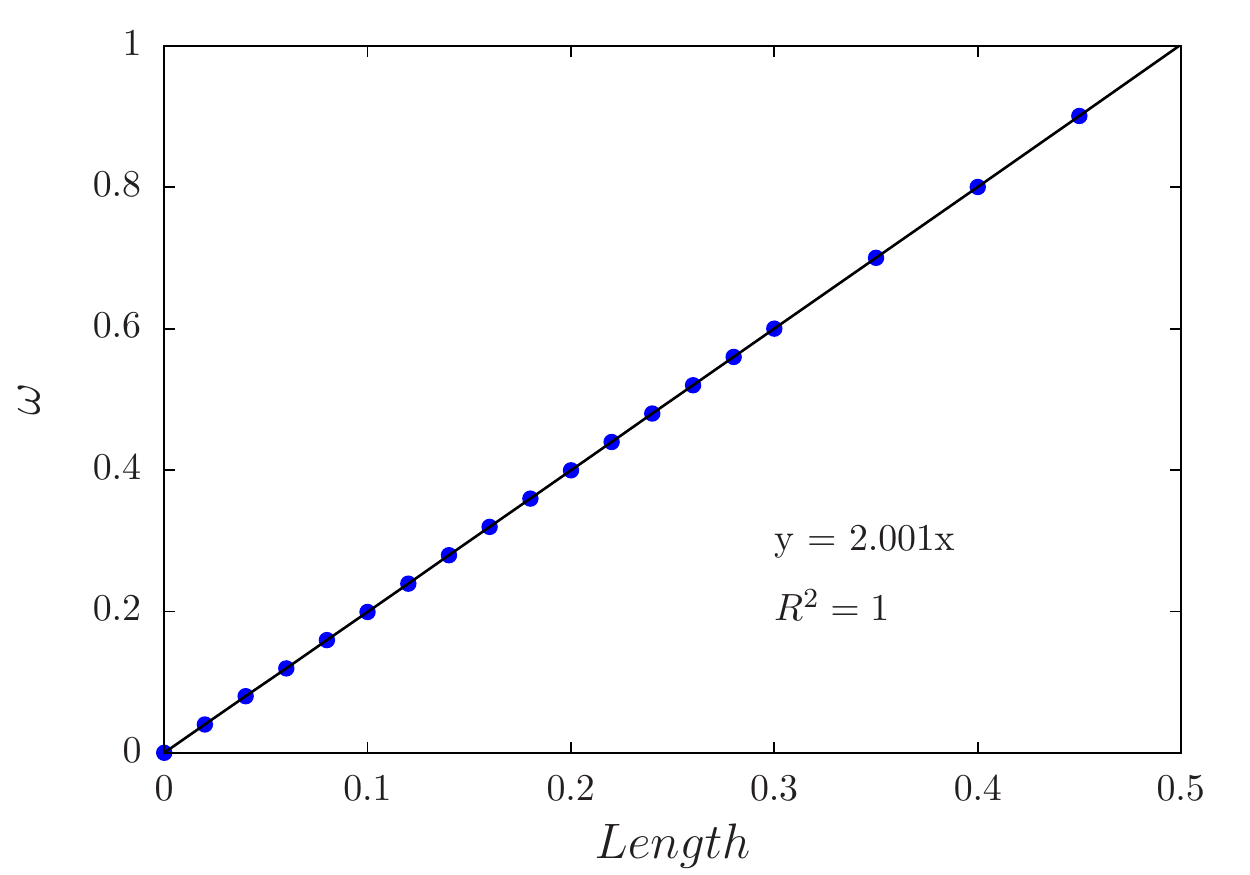}
\caption{Evolution of the capture area of a single finger with its
  length. Simulations show that the capture area of a finger that is
  not competing with any other neighbor is always twice its
  length.}\label{fig:single}
\end{figure}

Figure~\ref{fig:single} shows the evolution of the capture area
$\omega$ of a finger that is not in a competition regime. The capture
area grows linearly with the finger length.

\begin{figure}
\centering
\includegraphics[width=0.8\textwidth]{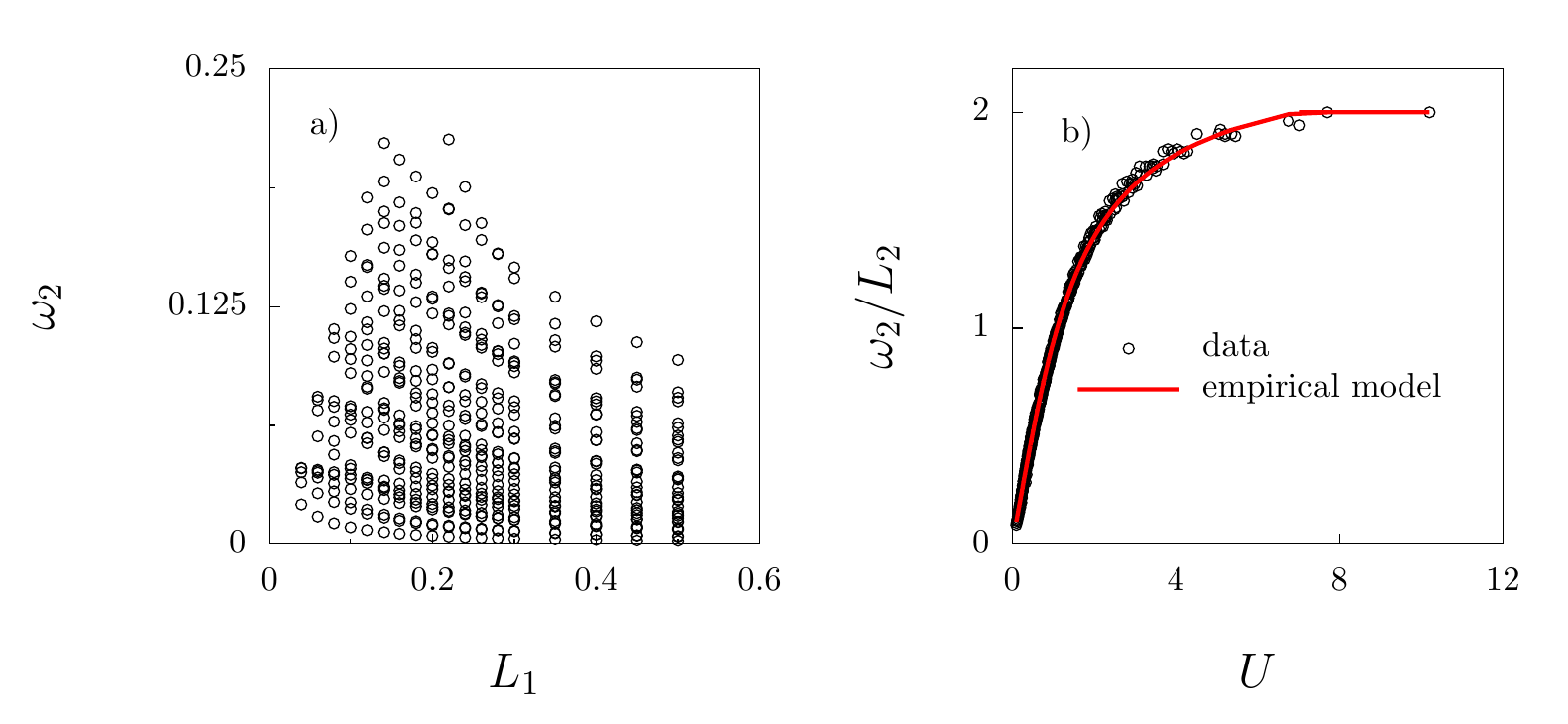}
\caption{(a) Capture area of the short finger ($\omega_{2}$) versus
  the length of the long one ($L_{1}$) in a two-finger system. It can
  be seen how the capture area of the short finger decreases as it is
  screened by the long one. (b) All the simulated points collapse when
  the capture area scaled by the finger length is plotted versus $U$
  (equation 11 of the manuscript).}\label{fig:fitting}
\end{figure}

Figure~\ref{fig:fitting} shows the comparison between the simulation
data and the empirical model for a two-finger system of longitudes
$L_{1}$ and $L_{2}$ with $L_{1}>L_{2}$ and variable distance $d$
between them. Figure~\ref{fig:fitting}a shows the dependence of the
short finger's capture area $\omega_{2}$ on the length of the the long
finger $L_{1}$. The data organizes itself into families characterized
by the distance between wormholes, which suggests that $\omega_{2}$
does not depend solely on $L_{1}$.

Figure~\ref{fig:fitting}b shows that the proposed model reproduces
$\omega_{2}$ satisfactorily for all tested $L_{1}$ and $d$. The
relationship between the fingers geometries is determinant for the result
of the competition. The short finger is heavily screened when the long
one is very near it (i.e., when $d \ll L_{1}$). Although the reduction in capture area decreases
as the lengths of the fingers become comparable.

%
%
\section{Concentration distribution along a wormhole}
The steady-state advection-dispersion-reaction equation in a wormhole
is
\begin{align}
  \label{eq:transport}
v(x)\frac{\partial c}{\partial x}= D \frac{\partial^2 c}{\partial y^2}
\end{align}
\begin{align*} 
  &0 \leq x \leq L\\
  &0 \leq y \leq b(x)
\end{align*}
where $x$ is the direction of the wormhole growth, $y$ is the
direction on the wormhole width, $c$ is concentration, $v(x)$ is the
velocity on the $x$ direction, $D$ is the diffusion coefficient and
$b$ is the half width in the $y$ direction.
Chemical reactions at the wall can be considered by imposing
\begin{align}
  \label{eq:bc-walls}
  D \nabla c \cdot \mathbf{n} \left.\right|_{y=b(x)}= -k_{f}c(x,b(x)).
\end{align}
where $k_{f}$ the kinetic constant and $c$ is the distance to the
equilibrium concentration. Under the assumption of nearly straight
wormholes, the width variation along $x$ can be neglected and the
boundary conditions written as
\begin{align}
  \label{eq:adv-disp_bound}
\left. D\frac{\partial c}{\partial y} \right.& \left.\right|_{y=b(x)} = -k_{f}c(x,b(x)) \\
\left. D\frac{\partial c}{\partial y}\right. & \left.\right|_{y=0} = 0\\
\label{eq:Dirichlet}
c(0,y)& = 1.
\end{align}

To solve \eqref{eq:transport} we perform two changes of variable. With
the first change of variable
\begin{align}
  \label{eq:y2z}
z=\frac{y}{b(x)}
\end{align}
we homogenize the wormhole width, converting the wormhole in a tube
with constant width. Concentration is written now as:
\begin{align}
  \label{eq:c_hat}
  c=\hat{c}(x,z(x,y))
\end{align}
and equation \eqref{eq:transport} takes the form
\begin{align}
  \label{eq:with_log}
  -v(x)\left[\frac{\partial \hat{c}}{\partial x}-z(x,y) \frac{d \log b(x)}{dx} \frac{\partial \hat{c}}{\partial z}\right] = \frac{D}{b^2(x)} \frac{\partial^2 \hat{c}}{\partial z^2}
\end{align}
Neglecting again the variation of $\log b(x)$ along the wormhole
length, \eqref{eq:with_log} can be written as
\begin{align}
  \label{eq:without_log}
  -\frac{b^2(x) v(x)}{D} \frac{\partial \hat{c}}{\partial x} = \frac{\partial^2 \hat{c}}{\partial z^2}
\end{align}
and the boundary conditions
\begin{align}
  \label{eq:boundary_1}
\left. \frac{D}{b(x)}\frac{\partial \tilde{c}}{\partial z} \right.& \left.\right|_{z=1} = -k_{f}\tilde{c} \\
\left. \frac{D}{b(x)}\frac{\partial \tilde{c}}{\partial z}\right. & \left.\right|_{z=0} = 0\\
\tilde{c}(0,z)& = 1.
\end{align}

We make now a second change of variables by defining
\begin{align}
  \label{eq:c_tilde}
  \tilde{c} (\tau(x),z) = \hat{c} (x,z) 
\end{align}
which from \eqref{eq:without_log} gives
\begin{align}
  \label{eq:with_tau}
\frac{b^2(x) v(x)}{D} \frac{\partial \tau}{\partial x} \frac{\partial \tilde{c}}{\partial \tau} = \frac{\partial^2 \tilde{c}}{\partial z^2}
\end{align} 

We define $\tau$ so that
$ \frac{b^2(x) v(x)}{D} \frac{\partial \tau}{\partial x}=1$, therefore
\begin{align}
  \label{eq:tau}
\tau= \int_{0}^{x} \frac{D}{b^2(x) v(x)} dx =\frac{D}{\omega q_{N}} \int_{0}^{x} \frac{dx}{b(x)},
\end{align}
so that \eqref{eq:with_tau} becomes
\begin{align}
  \label{eq:equation}
\frac{\partial \tilde{c}}{\partial \tau} = \frac{\partial^2 \tilde{c}}{\partial z^2}
\end{align}
with boundary conditions
\begin{align}
  \label{eq:boundary_2}
\left. \frac{D}{b}\frac{\partial \tilde{c}}{\partial z} \right.& \left.\right|_{z=1} = -k_{f}\tilde{c} \\
\left. \frac{D}{b}\frac{\partial \tilde{c}}{\partial z}\right. & \left.\right|_{z=0} = 0\\
\tilde{c}(0,z)& = 1.
\end{align}

We can solve \eqref{eq:equation} with the separation of variables
method, which gives
\begin{align}
  \label{eq:solution}
\tilde{c} (\tau,z) = \sum_{i=1}^{\infty} A_{i} \exp(-\lambda_{i}^{2}\tau) \cos( \lambda_{i}  z)
\end{align}
where $\lambda_{i}$ values are calculated from the boundary condition
at $z=1$~\eqref{eq:boundary_2}. At that boundary
\begin{align}
  \label{eq:solution_bound}
A_{i} \exp(-\lambda_{i}^{2}\tau) (-\sin( \lambda_{i} z) = -\frac{k_{f}b}{D} A_{i} \exp(-\lambda_{i}^{2}\tau) \cos( \lambda_{i} z).
\end{align}
Therefore $\lambda_{i}$ are the zeros of
\begin{align}
  \label{eq:lambda}
\lambda\tan(\lambda) = \frac{k_{f}b}{D}
\end{align}
which we computed by Newton's method.  

Finally, the values for the integration constants $A_{i}$ are obtained
from the inflow $(\tau=0)$ boundary condition by imposing (least
squares)
\begin{align}
  \label{eq:int:constants}
\tilde{c}_{0} = \sum_{i=1}^{\infty} A_{i} \cos( \lambda_{i} z)
\end{align}
%
%
\section{Dissolution at the wormhole walls and at the tip }
Using $x, y$ variables the length increment during a time step
$\Delta t$ of a wormhole should be given by the difference between the
total capacity of dissolution of the incoming water minus the capacity
spent in dissolving the wormhole walls. Actually, only a fraction
$\lambda$ of this capacity is spent in the wormhole advance. As we
will see in Section 4 below, a $\left(1-\lambda\right)$ fraction is
spent in the dissolution halo. Therefore,
\begin{align}
\Delta L = \lambda \left(q_{N} \omega - 2\int_{0}^{L} k_{f}c(x,b(x)) \, dx\right) \frac{v_{d}}{2b(L)} = \frac{\omega q_{N}\overline{c}}{b}
\end{align}
where $v_{d} = V_{M}(C_{0}-C_{eq})\Delta t /(1-\phi)$, with $V_{M}$
the molar volume and $\phi$ the matrix porosity, and it is assumed
that the wormhole grows with a width equal to the one at the tip
$2b(L)$.

This expression can be transformed into the $(\tau, z)$ space by using
\eqref{eq:y2z} and \eqref{eq:tau}, which gives
\begin{align}
\label{eq:deltaL-tau1}
\Delta L = \lambda \left(q_{N} \omega  - 2\int_{0}^{\tau(L)} k_{f}\frac{q_{N}\omega b(\tau)}{D}\hat{c}(\tau,1) \, d\tau\right) \frac{v_{d}}{2b(\tau(L))}.
\end{align}
We can further simplify \eqref{eq:deltaL-tau1} as
\begin{align}
\label{eq:deltaL-tau2}
 \Delta L = \lambda q_{N} \omega \left(\frac{1}{2} - \int_{0}^{\tau(L)} k_{f}\frac{b(\tau)}{D}\hat{c}(\tau,1) \, d\tau\right) \frac{v_{d}}{b(\tau(L))},
\end{align}
We can define the Damk{\"o}hler number as
\begin{align}
\label{eq:Damkohler}
\text{Da}^{\ast}(\tau) = \frac{k_{f}b(\tau)}{D}
\end{align}
so that
\begin{align}
\label{eq:deltaL-tau3}
 \Delta L = \lambda q_{N} \omega \left(\frac{1}{2} - \int_{0}^{\tau(L)}  \text{Da}^{\ast}(\tau) \hat{c}(\tau,1) \, d\tau\right) \frac{v_{d}}{b(\tau(L))}.
\end{align}

For the straight wormhole we use in the analytical solution,
\eqref{eq:tau} yields $\tau(L) = DL/\omega q_{N}b$ or, for a single
wormhole, $\tau = D/2q_{N}b$ since $\omega = 2L$. Note that
$\tau = 1/\text{Pe}$. Therefore \eqref{eq:deltaL-tau3} simplifies to
\begin{align}
\label{eq:deltaL-tau4}
 \Delta L = \lambda q_{N} \omega \left(1 - 2\text{Da}^{\ast} \int_{0}^{1/\text{Pe}}   \hat{c}(\tau,1) \, d\tau\right) \frac{v_{d}}{b}
\end{align}
and the solution only depends on $\text{Da}^{\ast}$, except for
$\tau=1/\text{Pe} = q_{N}\omega b/D$.

In general not all the dissolution capacity is spent in the advance of
the wormhole but also in the dissolution halo around the tip. This can
be accounted by multiplying \eqref{eq:deltaL-tau3} or
\eqref{eq:deltaL-tau4} by a factor $\lambda$, whose principal effect
is to modify the time scale of the wormhole advance.
%
%
\section{The dissolution halo surrounding wormholes}
\begin{figure}
\centering
\includegraphics[width=0.65\textwidth]{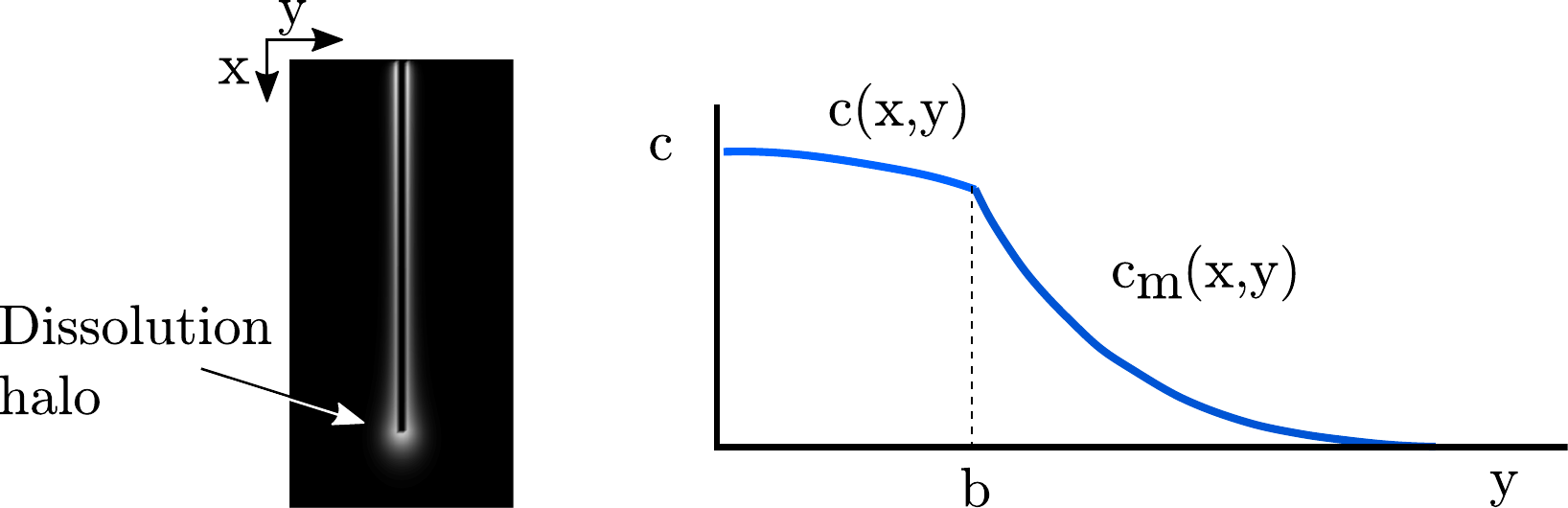}
\caption{Wormholes are surrounded by a dissolution halo (left). Beyond
  the wormhole wall, concentration decays exponentially with depth
  into the matrix $(y-b)$ and thickness characterized by
  $\ell_{m}$ (right).}\label{fig:halo}
\end{figure}

The boundary condition (2) is a simplification of what occurs in
reality. Dissolution does not take place at the wormhole wall, but
within the porous medium, at a dissolution halo surrounding the all
the wormhole (Figure~\ref{fig:halo}). As the wormhole advances the
halo at the wormhole's tip becomes part of the wormhole and reproduces
itself ahead of the wormhole. The presence of this halo raises two
questions regarding the validity of \eqref{eq:transport} --
\eqref{eq:Dirichlet} and the competition model to represent wormhole
growth. First, whether the thickness of the halo is too large (in
which case one cannot speak of wormholes) and second whether a high
transmissivity zone affects the flow distribution around the
wormhole. Prior to responding to these questions, we need to address
the concentration field around a wormhole.
%
%
\subsection{Concentration around a wormhole}
\begin{figure}
  \centering
  \includegraphics[width=0.9\textwidth]{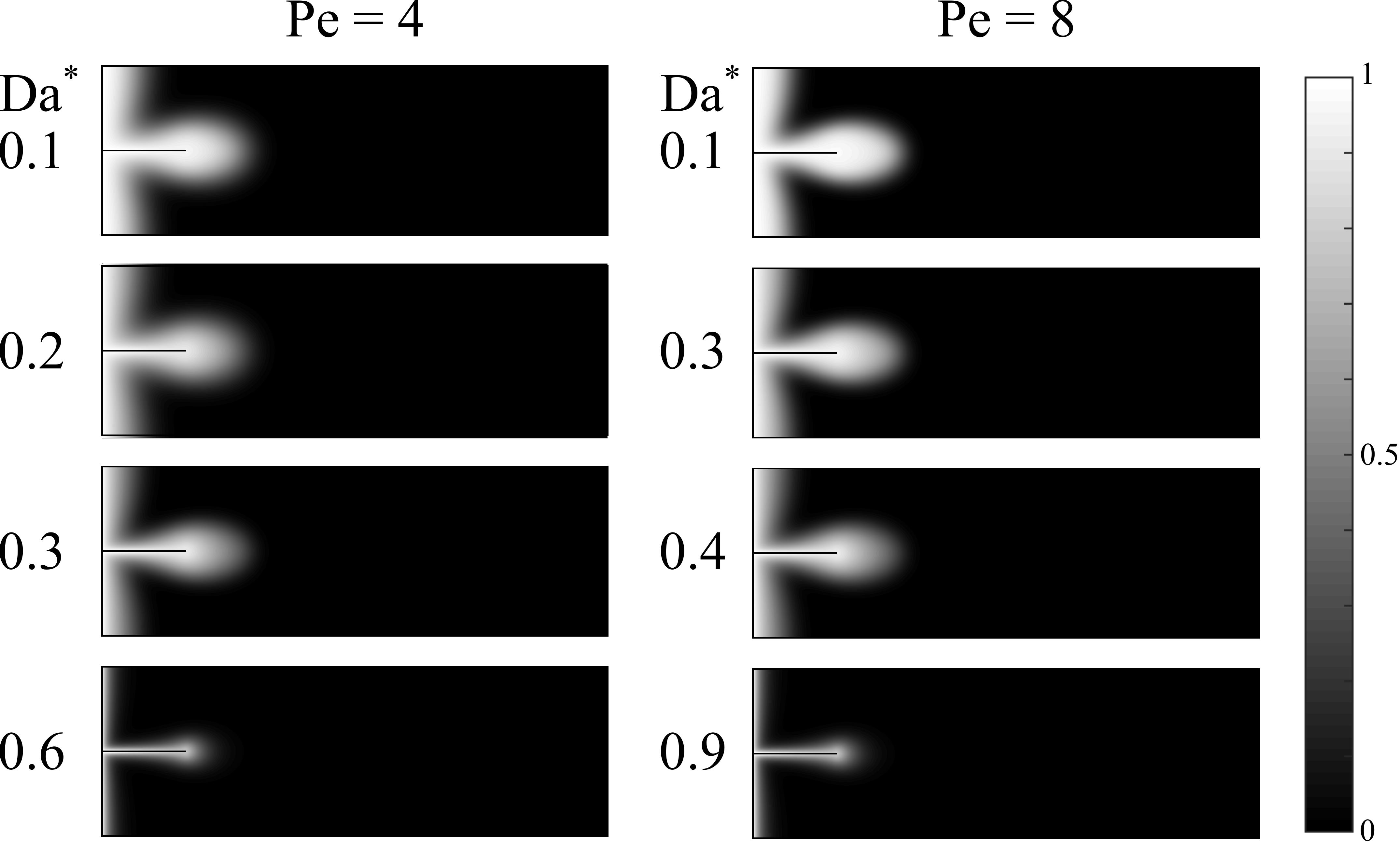}
  \caption{Concentration distribution around a single wormhole for
    several values of Damk{\"o}hler and P{\'e}clet numbers. Low
    $\text{Da}^{\ast}$ leads to uniform dissolution, that is,
    wormholes do not develop. As $\text{Da}^{\ast}$ increases the
    thickness of the halo reduces and the dissolution concentrates at
    the wormhole's tip.}\label{fig:halos}
\end{figure}

We simulate reactive transport around a wormhole, governed by
\begin{align}
  \label{eq:tpt}
  \phi_{m}\frac{\partial c_{m}}{\partial t}=D_{m}{\nabla^{2}c_{m}}- q_{m}\nabla{c_{m}}+ k_{m} \sigma c_{m}
\end{align}
where $D_{m}$ is the diffusion coefficient in the porous medium (equal
to $D$ multiplied by porosity and constrictivity and divided by
tortuosity squared), $c_{m}$ is the concentration within the porous
matrix, $q_{m}$ is water flux, $k_{m}$ is the reaction rate constant,
and $\sigma$ is the specific surface.  We have solved \eqref{eq:tpt}
numerically using compact finite differences, which is appropriate for
the type of instabilities leading to wormholes~\citep{Lele1992}. To
this end, we first solve the flow equation in a $3 \times 1$
rectangular domain, by imposing no flow at $x=-0.5$ and $x=0.5$, and
prescribing heads at $y=0$ and flux, $q_N$ at $y=3$. We simulate a
wormhole of half-width $b=0.01$, and length $L=0.5$ by assuming that
at the wormhole hydraulic conductivity is 100 times that of the matrix
and $k_m=0$.  Results for several values of Damk{\"o}hler and two
values of P{\'e}clet number are shown in
Figure~\ref{fig:halos}. Comparing the corresponding values of
Damk{\"o}hler and P{\'e}clet number to those in Figure 2 of the
manuscript, it is apparent that the reaction rate for the low
Damk{\"o}hler number would lead to uniform dissolution. That is, when
the reaction rate is too slow, wormholes do not develop because
dissolution occurs over a long distance within the porous matrix. Only
the largest Damk{\"o}hler number would lead to stable wormholes. We
have not simulated even larger reaction rates because, a very large
refinement would have been need for accuracy, but it can be
extrapolated that it would look like a narrow concentration halo
around the wormhole, with most dissolution occurring at the entrance,
except for extremely large P{\'e}clet, which would also be challenging
from a numerical point of view. An implication from this figure is
that dissolution does not occur at the walls but at a dissolution halo
around the wormhole.
%
%
\subsection{The halo thickness}
To address the question regarding the thickness of the halo, we solve
the reactive advection-diffusion problem in the matrix along the
streamlines while neglecting lateral gradients (from an analysis of
the last row of Figure~\ref{fig:halos} above and Figure~1 of the paper
it becomes apparent that flow occurs along the $y$ direction close to
the wormhole). Therefore we write
\begin{align}
  D_{m}\frac{\partial^{2}c_{m}}{\partial y^{2}} - q_{m}\frac{\partial c_{m}}{\partial y}= k_{m} \sigma c_{m}
\end{align}
The solution must be continuous at the wormhole wall $y = b(x)$ and to
vanish far away
\begin{align}
   \label{eq:continuity2}
c_{m}(x,b) = c(x,b)
\\
 \lim_{y\to \infty}c_{m}(x,y) = 0
\end{align}
Defining
\begin{align}
  \label{eq:ell}
  \ell_{m} = \frac{2D_{m}}{\sqrt{{q_{m}}^{2}+4D_{m}k_{m}\sigma}- q_{m}}
\end{align}
with $\ell_{m} = q_{m}/k_{m}\sigma$ when $D_{m}=0$, the solution to
this problem is well known
\begin{align}
  \label{eq:cm-solution}
  c_{m}(x,y) = c(x,b)e^{-(y-b)/\ell_{m}}
\end{align}
We use this solution to obtain the halo thickness but also the
equivalent surface reaction rate. To this end, we need to realize that
lateral dissolution occurs both by diffusion across the wall along the
whole wormhole but also by advection near the tip. To properly
separate the two effects, we define $k_{f}$ to represent dissolution
by diffusion. This is consistent with the solution in Section 2 of the
paper, where flow rate is constant along the whole wormhole. The
effect of advection on the development of the tip will be addressed
later. Therefore, neglecting $q_{m}$, we derive $k_{f}$ by imposing
\begin{align}
   \label{eq:continuity-derivative}
   D_{m}\frac{\partial c_{m}}{\partial y} (x,b) = D\frac{\partial c}{\partial y}(x,b) = - k_{f} c(x,b)
\end{align}
And substituting \eqref{eq:cm-solution}:
\begin{align}
-\frac{D_{m}}{\ell_{m}}c_{m}(x,b) =  -k_{f} c(x,b)
\end{align}
and finally
\begin{align}
  \label{eq:kf-km}
  k_{f} = \sqrt{D_{m}k_{m}\sigma},
\end{align}
because $\ell_{m} = \sqrt{D_{m}/k_{m}\sigma}$ for $q_{m} =0$ in \eqref{eq:ell} and $c_{m}(x,b) = c(x,b)$.
Therefore the surface dissolution rate $k_{f}$ can be directly
obtained from the conventional volumetric dissolution rate $k_{m}$
using \eqref{eq:kf-km}.

A few remarks about the obtained solution. First, $D_{m}$ and $\sigma$
are assumed to be constant. In reality they will vary with $y$ as
porosity increases. Second, the effective value of $b$ will be
slightly larger than the actual half-width of the wormhole. Finally,
the thickness of the halo $\ell_{m}$~\eqref{eq:ell} can be written,
neglecting advection, as
\begin{align}
  \label{eq:ell-diff}
\ell_{m} = \sqrt{\frac{D_{m}}{k_{m}\sigma}} = \frac{b}{\alpha}\frac{1}{\text{Da}^{\ast}}
\end{align}
where $\alpha = D/D_{m}$ is the relation between diffusion in the
matrix and in the wormhole. A typical value of $\alpha \approx 10$ can
represent the impact of porosity, constrictivity and tortuosity on
diffusion. From~\eqref{eq:ell-diff}, we can see that the size of the
halo will be small except for very small $k_{f}$, in which case no
wormhole would have developed in the first place.

To study in detail this thickness, we write $\ell_{m}$ \eqref{eq:ell}
in terms of the P\'eclet and Damk{\"o}hler numbers of the matrix as
\begin{align}
  \label{eq:ell-matrix}
  \frac{1}{\ell_{m}} = \frac{\text{Pe}_{m}}{2L}\left(\sqrt{1 + 4\frac{\text{Da}_{m}}{\text{Pe}^{2}_{m}}} - 1 \right) ,
\end{align}
where
\begin{align}
  \text{Pe}_{m} &= \frac{q_{m}L}{D_{m}}
\end{align}
and
\begin{align}
\text{Da}_{m} &= \frac{k_{m} \sigma L^{2}}{D_{m}}.
\end{align}

If the advection is not the dominant mechanism, that is
$\text{Da}_{m}/\text{Pe}^{2}_{m} \gg 1$, \eqref{eq:ell-matrix} can be
approximated as
\begin{align}
    \frac{1}{\ell_{m}} \approx \frac{\text{Pe}_{m}}{2L}\left(\sqrt{4\frac{\text{Da}_{m}}{\text{Pe}^{2}_{m}}} - 1 \right)  = \frac{1}{L} \sqrt{\text{Da}_{m}},
\end{align}
which is equivalent to~\eqref{eq:ell-diff} because~\eqref{eq:kf-km}
gives $\text{Da}^{\ast} \propto \sqrt{\text{Da}_{m}}$.

When advection dominates, that is,
$\text{Da}_{m}/\text{Pe}^{2}_{m} \ll 1$, we can write \eqref{eq:ell-matrix} as
\begin{align}
    \frac{1}{\ell_{m}} \approx \frac{1}{L}\frac{\text{Da}_{m}}{\text{Pe}_{m}},
\end{align}
which means that the size of the halo will be large if $q_{m}$ is
large. In such case, lateral wormholes may grow if $q_{m}$ is not very
large, which may explain why dominant wormholes often
bifurcate. Whether any of these bifurcations becomes dominant is a
problem similar to the one we are analyzing in this paper. However,
the main wormhole usually continues its advance leaving screened
wormholes as lateral dead branches \cite[see e.g. Figure 9c
in][]{Golfier2002}. The scenario in which $q_{m}$ is very large is not
possible because no wormholes would have developed in the first place
\cite[see e.g. Figure 9e in][]{Golfier2002}.

Finally, because of the lateral flow at the tip, part of the
dissolution capacity is used up in lateral dissolution (either
enlarging the dissolution halo or branching). This implies that no all
the dissolution capacity is used in advancing the wormhole. We term
$\lambda$ the fraction of the dissolution capacity at the tip that is
effectively used in the advancement.
\subsection{Impact of the high conductivity along the halo.}
\begin{figure}[htp]
  \centering
  \includegraphics[width=0.8\textwidth]{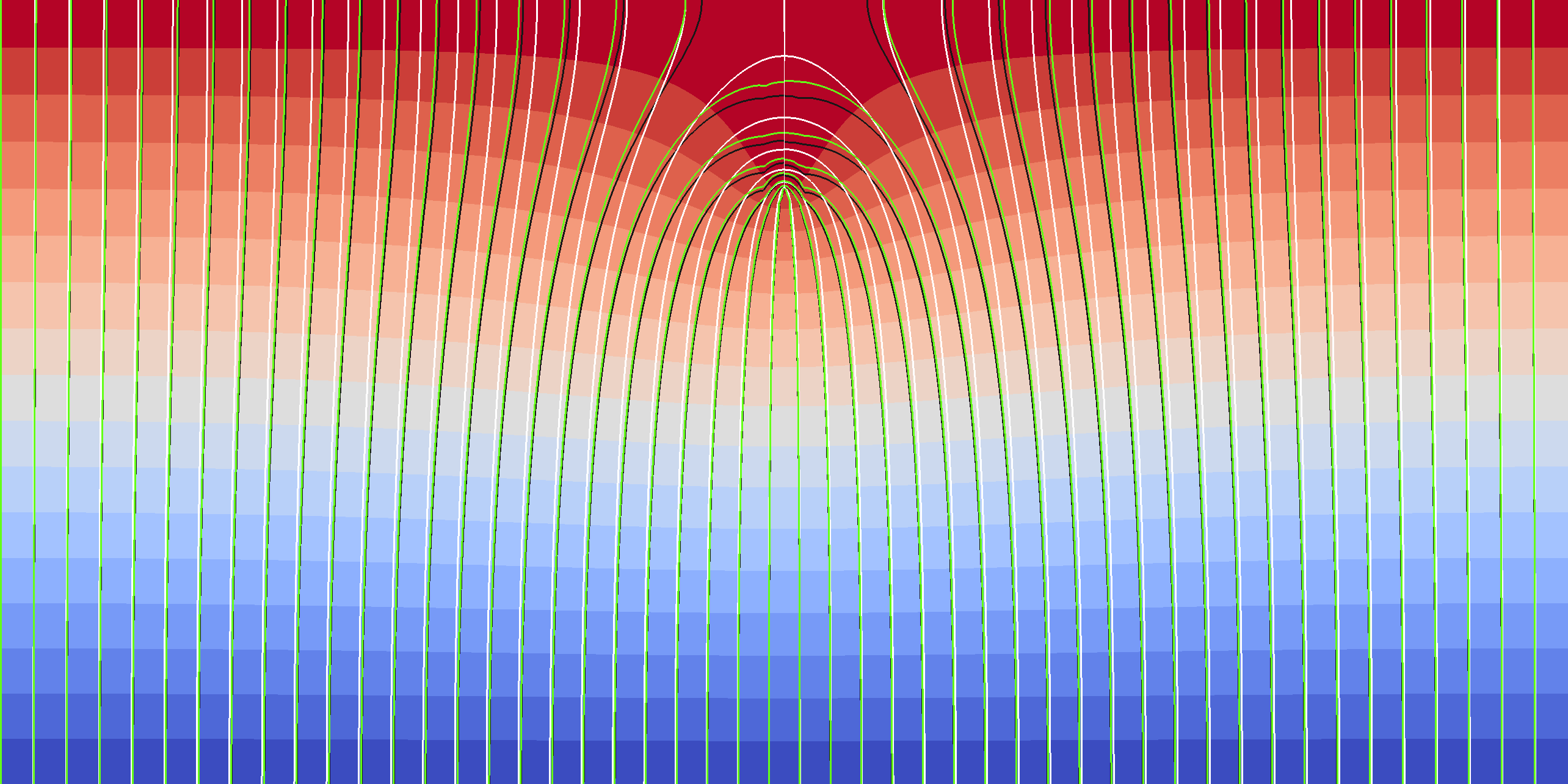}
  \caption{Streamlines followed by the water ejected by a wormhole for
    different transmissivity contrasts between the aquifer and the
    halo around the wormhole. White lines correspond to
    $T_{\text{halo}}/T_{\text{aq}} =1$, green lines to
    $T_{\text{halo}}/T_{\text{m}} =10$, and black
    $T_{\text{halo}}/T_{\text{aq}} =100$. The background color shows
    the head distribution for the case
    $T_{\text{halo}}/T_{\text{aq}} =100$.}\label{fig:streamlines}
\end{figure}

To test how the presence of the dissolution halo affects the flow
along the wormhole we performed a series of simulation in which the
halo was represented by a high transmissivity area around the
wormhole. Figure~\ref{fig:streamlines} shows the effect of a partially
dissolved region around the wormhole on the streamlines followed by
the water ejected from the wormhole. The simulations consider that the
transmissivity in a region around the wormhole has increased by a
factor of 1(white lines), 10 (green lines), and 100 (black lines). The
horizontal extension of the halo is 1/5 the length of the wormhole to
make sure that this effect is not underestimated.

\begin{figure}[htp]
  \centering
  \includegraphics[width=0.8\textwidth]{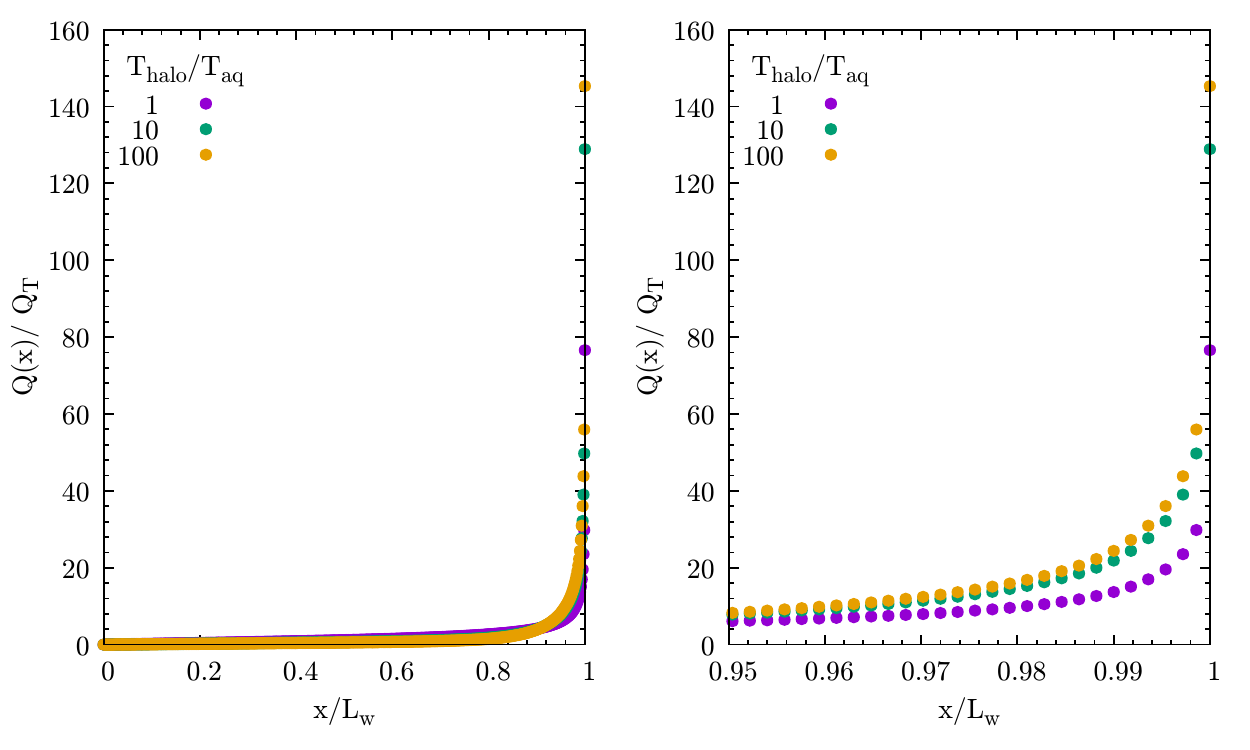}
  \caption{Flow distribution along a wormhole normalized by the total
    flow for different transmissivity contrasts between the matrix and
    the whormhole's surrounding area. The panel on the right shows the
    detail around the wormhole's tip.}\label{fig:flow}
\end{figure}

It can be seen that the extent of the capture area does not change
significantly. Increasing the transmissivity around the wormhole
causes the streamlines to be somewhat less concentrated around the tip
of the wormhole, but the effect is small. Figure~\ref{fig:flow} shows
how the flow concentrates at the tip of the wormhole for all cases,
which supports the hypothesis that the dissolution is controlled by
diffusion, i.e. $\text{Da}^{\ast}$, at the wormhole's walls and
concentrates at the wormhole's tip for the considered range of
$\text{Da}^{\ast}$. Note that the mesh discretization used for
Figure~\ref{fig:flow} was finer than for Figure~\ref{fig:streamlines}.
%
%

%